\documentclass[reprint,superscriptaddress,amsmath,amssymb,aps,a4paper,]{revtex4-2}

\usepackage{graphicx}% Include figure files
\usepackage{dcolumn}% Align table columns on decimal point
\usepackage{bm}% bold math
\usepackage{multirow}
\usepackage[mathlines]{lineno}% Enable numbering of text and display math
\usepackage[normalem]{ulem}% strikethrough
\usepackage{xcolor}% font colors
\begin{document}

\newcommand{\ccnumu}{CC~$\nu_\mu$}
\newcommand{\ccnue}{CC~$\nu_e$}
\newcommand{\valerr}[2]{#1$\,\pm\,$#2}

%\preprint{APS/123-QED}

\title{Application of Transfer Learning to Neutrino Interaction Classification}

%% EPJC format
% \author[1]{\fnm{Andrew} \sur{Chappell}}\email{andrew.chappell@warwick.ac.uk}

% \author[2]{\fnm{Leigh H.} \sur{Whitehead}}\email{leigh.howard.whitehead@cern.ch}
% %\equalcont{These authors contributed equally to this work.}

% \affil[1]{\orgdiv{Department of Physics}, \orgname{University of Warwick}, \orgaddress{\city{Coventry}, \postcode{CV4 7AL}, \country{United Kingdom}}}

% \affil[2]{\orgdiv{Department of Physics}, \orgname{University of Cambridge}, \orgaddress{\city{Cambridge}, \postcode{CB3 0HE}, \country{United Kingdom}}}
%% End EPJC

%% RevTex author things
\newcommand{\Cambridge}{Department of Physics, University of Cambridge, Cambridge, CB3 0HE, United Kingdom}
\newcommand{\Warwick}{Department of Physics, University of Warwick, Coventry, CV4 7AL, United Kingdom}
\affiliation{\Cambridge}
\affiliation{\Warwick}

\author{Andrew Chappell}
 \email[Email: ]{andrew.chappell@warwick.ac.uk}
\affiliation{\Warwick}

\author{Leigh H. Whitehead}
 \email[Email: ]{leigh.howard.whitehead@cern.ch}
\affiliation{\Cambridge}
%% End RevTex

\date{\today}% It is always \today, today,

%% EPJC
%\abstract{We will write an abstract here soon once we have done some studies and got some results}

%% RevTex format
\begin{abstract}
Training deep neural networks using simulations typically requires very large numbers of simulated events. This can be a large computational burden and a limitation in the performance of the deep learning algorithm when insufficient numbers of events can be produced. We investigate the use of transfer learning, where a set of simulated images are used to fine tune a model trained on generic image recognition tasks, to the specific use case of neutrino interaction classification in a liquid argon time projection chamber. A ResNet18, pre-trained on photographic images, was fine-tuned using simulated neutrino images and when trained with one hundred thousand training events reached an F1 score of \valerr{0.896}{0.002} compared to \valerr{0.836}{0.004} from a randomly-initialised network trained with the same training sample. The transfer-learned networks also demonstrate lower bias as a function of energy and more balanced performance across different interaction types.
\end{abstract}

                              
\maketitle

%\tableofcontents

\section{Introduction}\label{sec:introduction}
The usage of Deep Learning has increased rapidly in neutrino physics over the last five to ten years~\cite{Psihas2020,hepReview}. The data from many neutrino experiments can be easily and naturally represented in an image format, hence Convolutional Neural Networks (CNNs) are a very popular choice of deep learning algorithm in the field. CNN models contain millions of parameters that must be trained, which is typically done using large numbers of simulated neutrino interactions. For example, the CNN used to perform the neutrino event classification in the Deep Underground Neutrino Experiment (DUNE) was trained on over three million simulated events~\cite{DUNE:2020gpm,AlonsoMonsalve:2751646}. 

However, detector simulations for large detectors are very time consuming and resource intensive, so other methods are being explored to be able to train powerful and accurate deep learning algorithms without a very large computational burden. Potential solutions to this problem fall in to three categories: methods to make faster simulations, methods to improve computational performance of the networks and methods to reduce the number of simulated events required. However, the use of GPUs in deep learning can carry its own computational burden, and this resource intensity is becoming of increasing importance in the light of high energy costs and increased focus on the carbon footprint of research activities~\cite{climateAndHEP} and we must therefore ensure we use such resources as effectively and efficiently as possible. Methods to make faster simulations often use a generative model, typically a Generative Adversarial Network, to approximate the simulation to produce events much more quickly (see, for example, Chapter 6 of Ref.~\cite{hepReview2} for a review). To improve computational performance, alternative network architectures using sparse representations of the images have been deployed (see, for example,  Ref.~\cite{Domin2020}). For reduction of event requirements \textit{transfer learning} can be used as an approach to use a much smaller number of simulated events to fine tune an existing, pre-trained model, with these  models often trained on photographic images.

Transfer learning was first proposed in 1976 by Bozinovski and Fulgosi~\cite{transfer1,transfer2} for the training of perceptrons. More recently it has been applied to deep learning~\cite{transferReview}, including in fields similar to neutrino physics: an example from the AT-TPC nuclear physics experiment showed that a fairly small number (thousands) of training examples gave good performance when used to fine tune a generically pre-trained model~\cite{Kuchera:2018djs}.

As in the AT-TPC experiment, liquid argon time projection chamber (LArTPC) event displays bare little resemblance to the photographic images used to train existing models, and therefore the goal of this article is to assess the effectiveness of using transfer learning for the classification of interactions in a DUNE-like LArTPC detector and determine the most appropriate approaches to fine tuning. The details of the event simulation and image production are given in Sec.~\ref{sec:simulation}, Sec.~\ref{sec:classification} presents a case-study with the aim of classifying three general types of neutrino interactions, Sec.~\ref{sec:results} presents the results of the study, and Sec.~\ref{sec:discussion} provides a discussion and closing remarks.

\section{Simulated Event Samples}\label{sec:simulation}

Neutrino interactions were generated using GENIE \texttt{v3\_00\_06}~\cite{Alam:2015nkk} and a uniform flux distribution between 1\,GeV/$c^2$ and 4\,GeV/$c^2$. The flux distribution was chosen to give a rough approximation of the DUNE flux in the main oscillation region of the spectrum~\cite{DUNE:2020jqi}. Three balanced samples of interaction were produced: charged-current muon neutrino (\ccnumu{}), CC electron neutrino (\ccnue{}), and neutral current (NC) interactions. The important outputs from GENIE in this case are the kinematics of the incoming neutrino and the argon target, and the kinematics of all of the final-state particles produced in the interaction.

\begin{figure*}[htb]
    \centering
    \begin{tabular}{ccc}
    \includegraphics[width=0.32\textwidth]{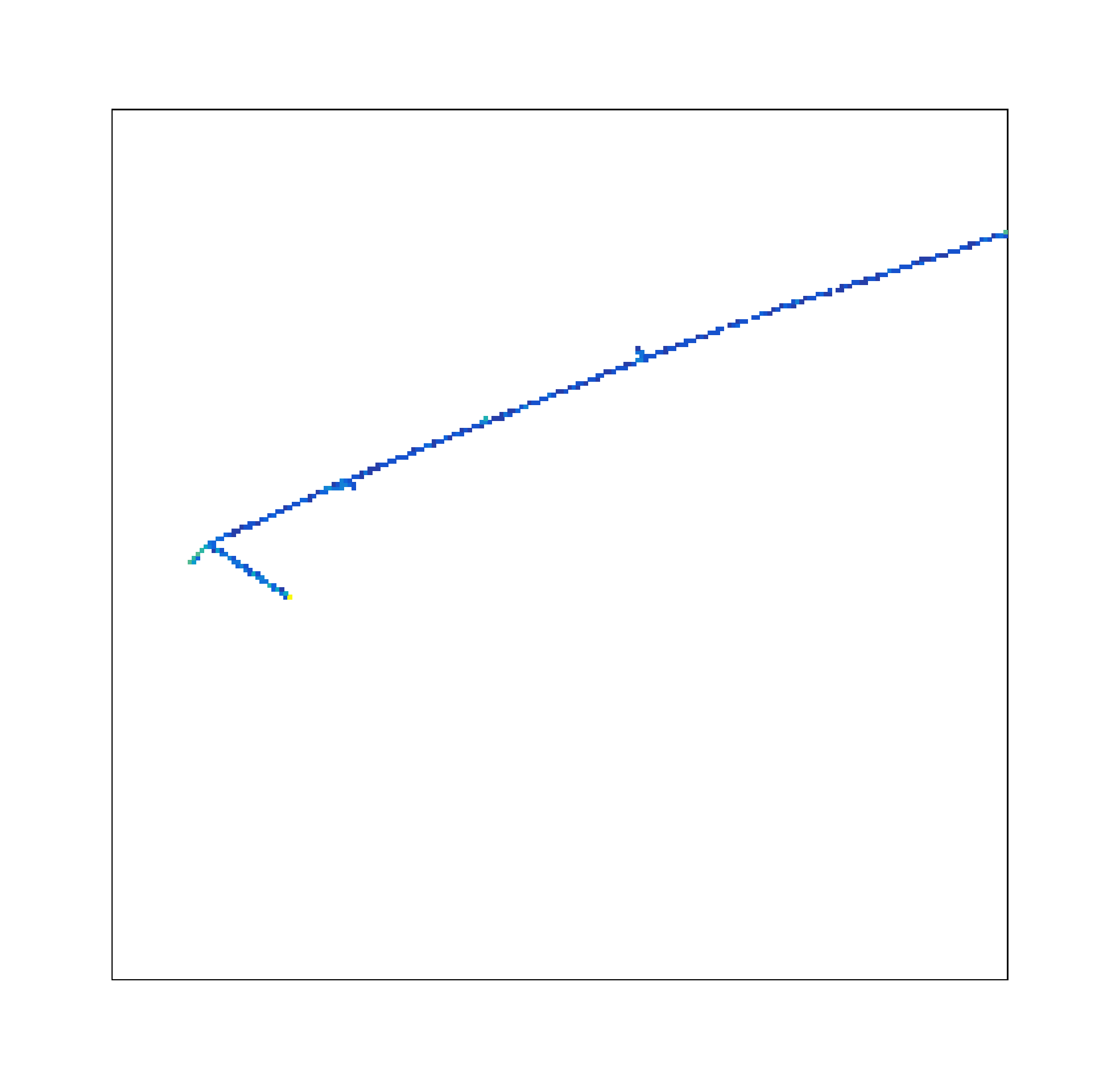}
    & \includegraphics[width=0.32\textwidth]{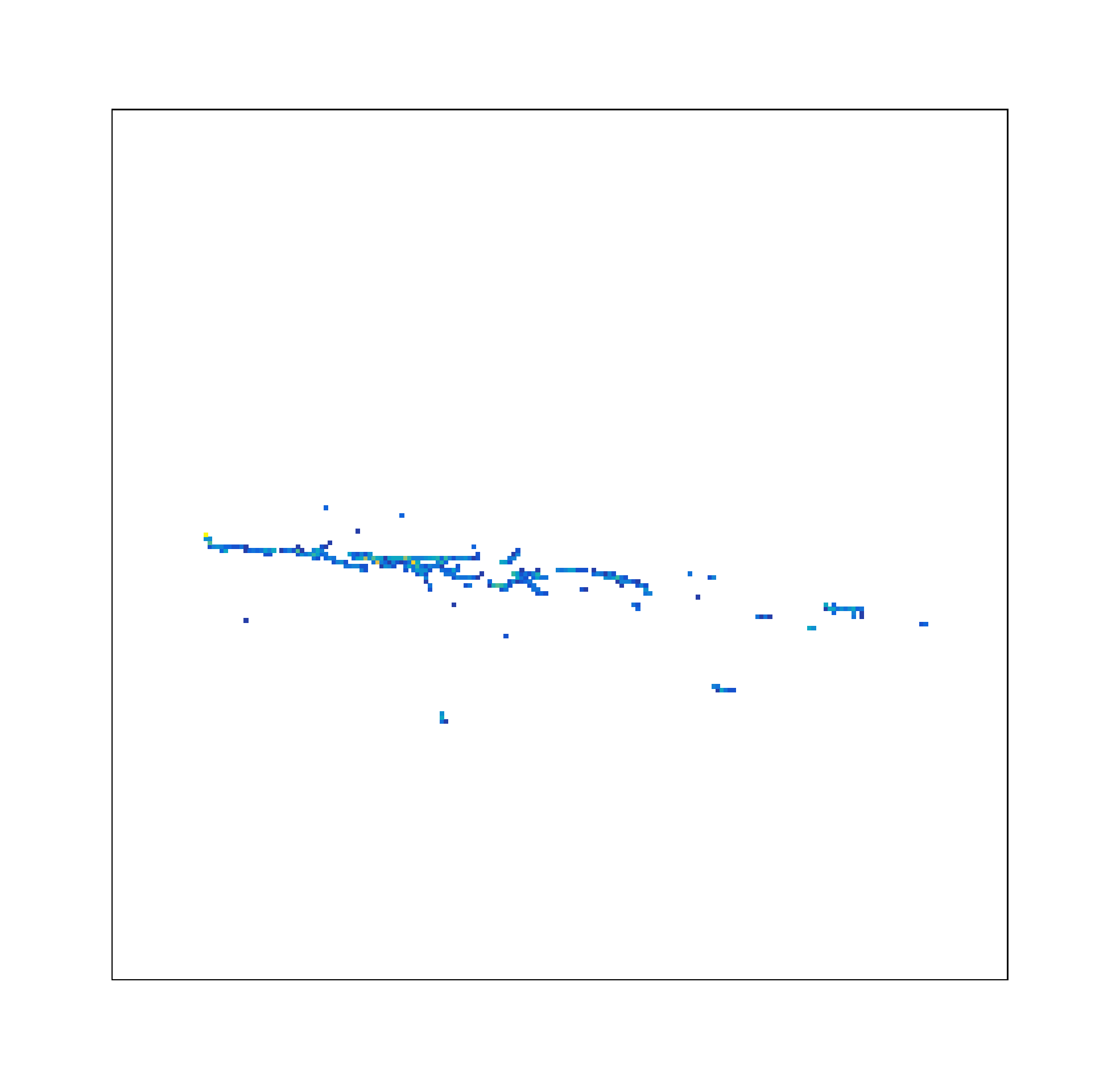}
    & \includegraphics[width=0.32\textwidth]{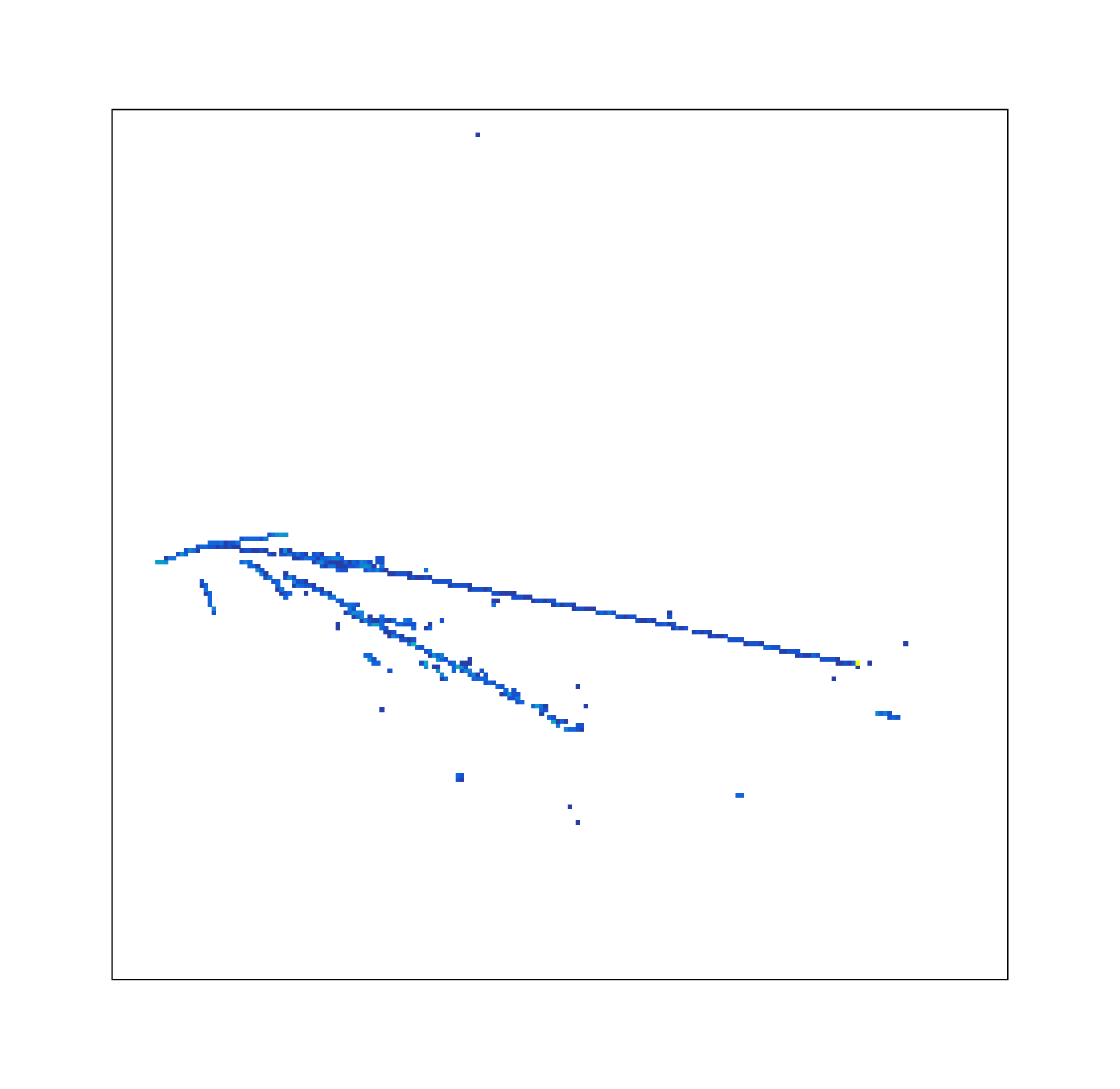}
    \end{tabular}
    \caption{Three simulated neutrino interactions shown in the $(w,x)$ view. Examples of a \ccnumu{}, a \ccnue{} and an NC interaction are shown in the left, centre and right panels, respectively. Each image is 224$\times$224 pixels, each pixel corresponds to an area of 1$\times$1\,cm$^2$, and the colour represents the deposited energy (blue is lowest, yellow is highest).}
    \label{fig:eventdisplays}
\end{figure*}

The final-state particles are tracked through a simple LArTPC detector using Geant4 \texttt{v4\_10\_6}~\cite{Agostinelli:2002hh}. The detector geometry is defined as a cuboid filled with liquid argon and the dimensions in the $(x,y,z)$ directions are $5\,$m$\,\times\, 5\,$m$\,\times\, 5\,$m, where $z$ defines the beam direction, $y$ is vertical and $x$ is the drift direction. The simulation produces three-dimensional energy deposits within the detector volume that are projected into three two-dimensional views of the $yz$ plane, similar to the three wire readout planes in the planned DUNE detectors~\cite{DUNE:2020txw}. These three views are referred to as $u$, $v$ and $w$ and are aligned at 35.9$^\circ$, -35.9$^\circ$ and 0$^\circ$ to the vertical, respectively.

The output of the simulation is formed by three two-dimensional images showing $u$, $v$, and $w$ on the horizontal axis, and the drift coordinate $x$ on the vertical axis. The pixel intensity is given by the amount of energy deposited. Examples of a \ccnumu{}, a \ccnue{} and an NC interaction are shown in the left, centre and right panels of Fig~\ref{fig:eventdisplays}, respectively. Each event is shown in the $(w,x)$ view. The \ccnumu{} event shows the characteristic long muon track, the \ccnue{} event shows the typical electron shower emanating from the interaction vertex, and the example NC event shows that NC events can sometimes include components similar to the \ccnumu{} and \ccnue{} interactions. The chosen pre-trained network requires 224 $\times$ 224 pixel input images, so the images are produced such that each pixel represents a 1$\times$1\,cm$^2$ region, cropped and centred on the region surrounding the interaction. This represents an approximately twofold decrease in resolution compared to the $\sim$5\,mm granularity of the readout planes in order to contain larger interactions within the images. These pre-trained networks are used to classify photographic images, hence the three images of each event are stacked together to produce a depth three image that is analogous to a colour image with red, green and blue colour channels.

The total number of images available was 140,000, of which 20,000 events were used as a validation set and another 20,000 images as a test set to produce the final results.

\section{Event Classification Case Study Overview}\label{sec:classification}

\begin{figure*}[tbh]
  \centering
  \includegraphics[width=0.85\textwidth]{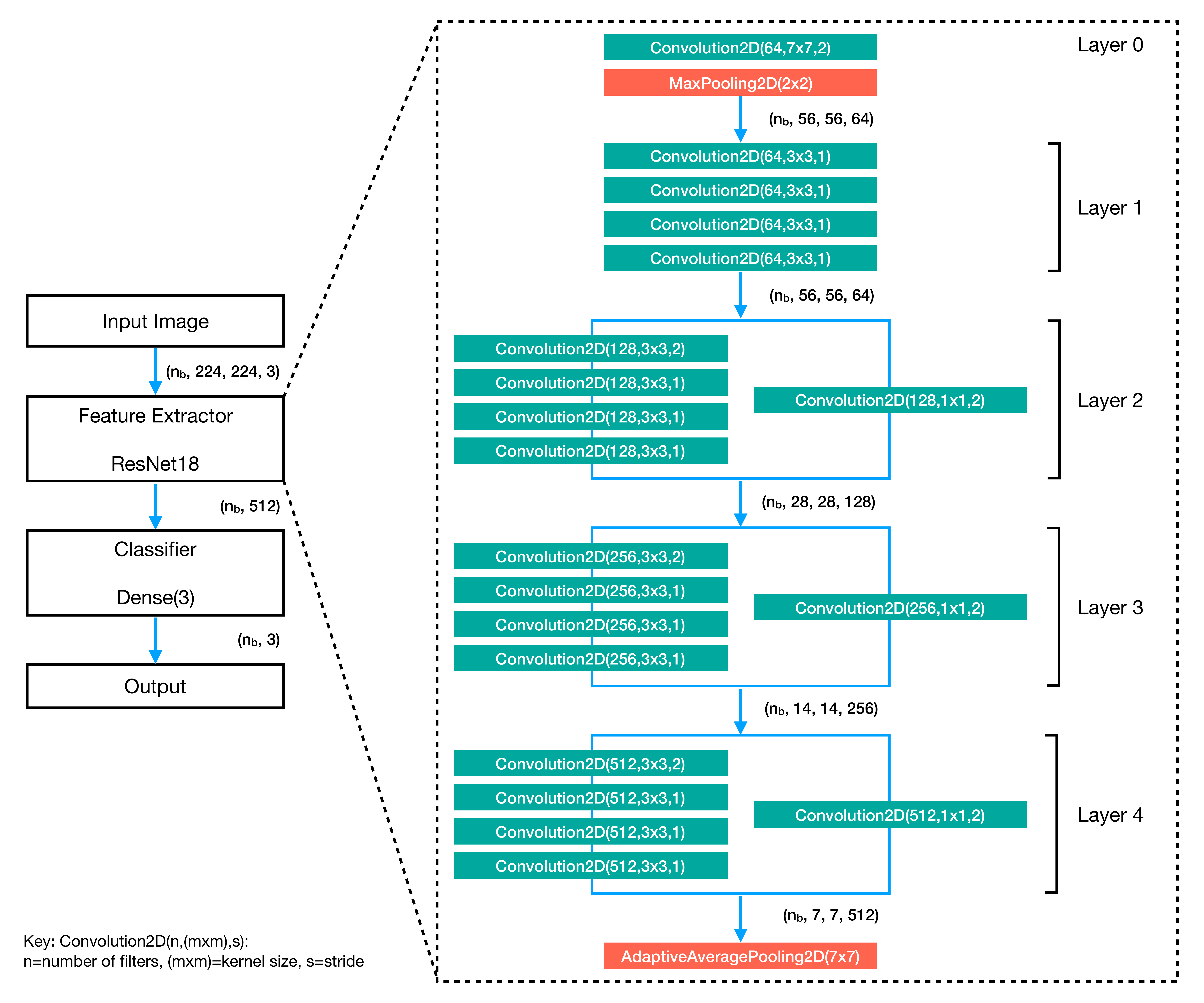}
  \caption{An overview of the architecture, where $224\times224$ pixel images of depth three are input to the ResNet18 feature extractor that outputs a 512 element tensor. The classifier takes the 512 element tensor and outputs three classification scores, one for each class. The number of images per batch, $n_b$, is shown in the tensor dimensions. For clarity, only the convolutional and pooling layers are shown for the feature extractor. The convolutional layers with a stride of two perform a downsampling of a factor of two in the image size, and the residual connections are those with the single 1$\times$1 convolutional layers. Following the ResNet convention, the convolutional layers are combined into groups called Layers 1 to 4.}
  \label{fig:simpleArchitecture}
\end{figure*}

The aim of this study is to investigate the use of transfer learning to train a CNN for the task of neutrino event classification. In the simplest case, long-baseline neutrino oscillation experiments need to be able to accurately and efficiently identify \ccnumu{}, \ccnue{}, and NC interactions. Each neutrino interaction will therefore be classified as one of the three true categories: \ccnumu{}, \ccnue{} or NC. 

The PyTorch~\cite{NEURIPS2019_9015} framework was used because of the wide range of pre-trained architectures available. The architecture chosen was the ResNet18~\cite{He-et-al-2015-deep}, since ResNets are a popular choice in the neutrino physics field and the relatively shallow depth eases the computational burden for training the hundreds of networks required by this study.

\subsection{Network Architecture}\label{sec:arch} 
We consider the architecture as two sub-networks: the \textit{feature extractor} network and the \textit{classifier} network. The feature extractor network consists of the many convolutional layers that extract features from the input images, and the classifier network, that provides the specific outputs for the task being performed. The classifier is specific to each use case, so must be appended to the predefined ResNet18 feature extractor network. The choice for the classifier was a single three node dense layer taking the ($n_b$, 512) output from the ResNet18 and returning the final three classification scores, where $n_b$ is the number of images per batch. The final architecture is shown in Fig.~\ref{fig:simpleArchitecture}, where only the convolutional and pooling layers of the ResNet18 have been shown for clarity. The naming convention from the ResNet architecture is used here, which breaks up the network into six blocks of layers, with the middle four blocks named Layer 1 to Layer 4. The total number of trainable weights in the network is 11,178,051, divided between the layers as follows: 9,536 (Layer 0); 147,968 (Layer 1); 525,568 (Layer 2); 2,099,712 (Layer 3); 8,393,728 (Layer 4); 1,539 (classifier).

\subsection{Training Details}
Stochastic gradient descent (SGD) was chosen as the optimiser. In all cases, the starting learning rate was set to 0.001, and it was reduced by a factor of 10 each time the validation loss did not improve for three epochs. The network training stopped automatically when the validation loss did not reduce for six epochs. A batch size of 32 images was chosen as an optimisation between classification performance, training time and memory usage. All networks were trained using a NVIDIA Tesla V100.

\subsection{Performance Metrics}
The F1 score~\cite{f1score} can be written in terms of true positives $\left(T_{p,i} \right)$, false positives $\left(F_{p,i} \right)$ and false negatives $\left(F_{n,i} \right)$, where the suffix $i$ indicates the target class under consideration.
%\begin{equation}\label{eq:f1score}
%    F_1 = \frac{T_p}{T_p + 0.5\left(F_p+F_n\right)}
%\end{equation}
For a multi-class classification with $n$ classes, such as in the analysis presented here with $n = 3$, the overall F1 score can be calculated from the individual scores for each class in a number of ways. Here the \textit{macro} averaging scheme is used, such that the metric is computed on a per-class basis and then the F1 score is taken to be the unweighted mean of the per-class scores:
%\begin{equation*}
%T_p = \sum_{i=1}^{3}T_{p,i}\,,~
%F_p = \sum_{i=1}^{3}F_{p,i} ~\textrm{and}~
%F_n = \sum_{i=1}^{3}F_{n,i}\,.
%\end{equation*}

\begin{equation}\label{eq:f1score}
    F_1 = \frac{1}{3}\sum_{i=1}^{3}\frac{T_{p,i}}{T_{p,i} + 0.5\left(F_{p,i}+F_{n,i}\right)}
\end{equation}

Equation~\ref{eq:f1score} shows that the allowed values of the F1 score are between zero and one, where one is the perfect score when there are no false positives or false negatives ($F_{p,i} = F_{n,i} = 0$). It also considers false positives and false negatives as equally bad in terms of calculating the score.

For each study presented, an ensemble of 25 independently trained versions of the network was produced, with the mean and error on the mean of the different metrics forming the reported results. The use of ensembles accounts for random fluctuations in the initialisation and training of the networks that arise from the fact that these are stochastic processes. It is important to note that when using a fixed random seed the results are deterministic and reproducible on a given system.

\section{Results}\label{sec:results}

\subsection{Randomly Initialised ResNet18}\label{sec:class:baseline}
ResNet18s with random weight initialisation form the baseline for this study against which the transfer learning results will be compared in Sec.~\ref{sec:final_results}. The Kaiming (also known as He)~\cite{kaiming} initialisation scheme was developed specifically for CNNs using non-linear activation functions such as ReLU. The specific version of the Kaiming initialisation scheme used in this work was the normal distribution form. The standard deviation of the distribution depends on the number of weights in the layer (equivalent to the size of the output from the previous layer), $n_w$: $\sigma = \sqrt{2 / n_w}$.

An ensemble of 25 networks were trained with differing numbers of training events used: 1,000 (1k); 2k; 3k; 5k; 7k; 10k; 15k; 20k; 30k; 40k; 50k; 75k; and 100k. 
Table~\ref{tab:riresults} shows the F1 scores from the testing and validation samples from the networks trained with the above number of interactions. As expected, the performance increases significantly as the number of training events rises. The uncertainty on the F1 score is seen to reduce as a function of the number of training images, which is expected as the training should become more stable to more training examples. Using the full training dataset of 100k events, an F1 score of \valerr{0.836}{0.004} was measured.

\begin{table*}[htb]
    \centering
    \caption{The F1 scores for the testing sample for the Kaiming initialised networks. The values and uncertainties given come from the mean and error on the mean of 25 different training attempts.}
    \vspace{10pt}

    \begin{tabular}{ccccc}%c}
        \hline \hline
         \noalign{\smallskip}
         %% Fiddling for EPJC         
         % Model & Initialisation & Training & Testing & Validation \\
         % & & Images & F1 Score & F1 Score \\
         %% RevTex
         ~Model~ & ~Initialisation~ & ~Training Images~ & ~Testing F1 Score~\\% & ~Validation F1 Score~\\
         \noalign{\smallskip}
         \hline 
         \noalign{\smallskip}
         & & 1k & \valerr{0.528}{0.014}\\% & \valerr{0.525}{0.013} \\
         & & 2k & \valerr{0.621}{0.010}\\% & \valerr{0.618}{0.010} \\
         & & 3k & \valerr{0.661}{0.010}\\% & \valerr{0.658}{0.010} \\
         & & 5k & \valerr{0.720}{0.010}\\% & \valerr{0.720}{0.010} \\
         & & 7k & \valerr{0.747}{0.009}\\% & \valerr{0.748}{0.010} \\
         & & 10k & \valerr{0.770}{0.008}\\% & \valerr{0.772}{0.008} \\
         ResNet18 & Kaiming & 15k & \valerr{0.787}{0.008}\\% & \valerr{0.788}{0.009} \\
         & & 20k & \valerr{0.800}{0.004}\\% & \valerr{0.803}{0.004} \\
         & & 30k & \valerr{0.816}{0.005}\\% & \valerr{0.819}{0.005} \\
         & & 40k & \valerr{0.816}{0.004}\\% & \valerr{0.819}{0.004} \\
         & & 50k & \valerr{0.822}{0.005}\\% & \valerr{0.824}{0.005} \\
         & & 75k & \valerr{0.833}{0.004}\\% & \valerr{0.835}{0.004} \\
         & & 100k & \valerr{0.836}{0.004}\\% & \valerr{0.838}{0.004} \\
         \noalign{\smallskip}
         \hline \hline
    \end{tabular}
    \label{tab:riresults}
\end{table*}

\subsection{Transfer Learning with ResNet18}\label{sec:class:tf}
The pre-trained ResNet18 that forms the basis of this study was trained on the ImageNet~\cite{imagenet_cvpr09} data sample, meaning that it was trained on photographic images with the goal of classifying them into one of one thousand categories. The classifier network was modified to provide the three required outputs for this use case, as described in Sec.~\ref{sec:arch}.

Samples of neutrino interaction images were then used to fine tune the weights of the pre-trained networks. Different networks have been trained with different numbers of training images, ranging from one thousand to one hundred thousand (with approximately equal fractions from each of the three true classes). The performance has been studied as a function of the number of ResNet18 Layers (as defined in Fig.~\ref{fig:simpleArchitecture}) with weights that are allowed to be fine-tuned, where the weights of the ResNet18 layers were progressively frozen:
\begin{itemize}
    \item \emph{All Weights}: No weights frozen, total of 11,178,051 trainable weights.
    \item \emph{Freeze(1)}: Layer 0 weights frozen, total of 11,168,515 trainable weights remain.
    \item \emph{Freeze(2)}: Layer 0 and 1 weights frozen, total of 11,020,547 trainable weights remain.
    \item \emph{Freeze(3)}: Layer 0, 1 and 2 weights frozen, total of 10,494,979 trainable weights remain.
    \item \emph{Freeze(4)}: Layer 0, 1, 2 and 3 weights frozen, total of 8,395,267 trainable weights remain.
    \item \emph{Classifier Only}: All ResNet18 weights frozen, total of 1,539 trainable weights remain.
\end{itemize}
When using a pre-trained feature extractor, it is clear that the classifier weights are the only ones that \textit{must} be trained in order to get performance better than random guessing. Beyond this, training Layer 4 will likely give the biggest step in performance because it contains approximately 75\% of the network weights. Generally, it is expected that the performance will improve as the number of trainable parameters increases. Furthermore, of interest to this use case is the evolution of the features that can be extracted at each layer of the network. %\citeauthor{zeiler}~\cite{zeiler} 
Zeiler and Fergus~\cite{zeiler} showed that early CNN layers comprise low-level geometric features (edges, corners, etc), with deeper layers becoming increasingly more class-specific. The expectation therefore, is that early layers may retain a high degree of relevance when applied to this use case, while deeper layers will increasingly contain many feature extractors of little relevance to our use case. One might therefore expect that fine tuning can be limited to deeper layers, which an assessment of performance by layer will also determine.

\subsection{Comparison of Transfer Learning Cases}

\begin{table*}%[htb]
    \centering
    \caption{The testing F1 scores for various combinations of the transfer learning cases and number of training images. The reported values show the mean value of the 25 trained networks and the associated error on the mean.}
    \vspace{10pt}
    \begin{tabular}{ccccc}%c}
         \hline \hline
         \noalign{\smallskip}
%% Fiddling for EPJC         
        Model & Initialisation & Weights & Training & Testing  \\
         & & Scheme & Images & F1 Score \\
%% RevTex
         %~Model~ & ~Initialisation~ & ~Weights Scheme~ & ~Training Images~ & ~Testing F1 Score~ \\%& ~Validation F1 Score~\\
         \noalign{\smallskip}
         \hline
         \noalign{\smallskip}
         & & All Weights & 100k & \valerr{0.896}{0.002}\\%& \valerr{0.895}{0.002} \\
         & & Freeze(1) & 100k & \valerr{0.898}{0.001}\\% & \valerr{0.897}{0.001} \\
         ResNet18 & Pre-trained & Freeze(2) & 100k & \valerr{0.892}{0.002}\\% & \valerr{0.892}{0.002} \\
         & & Freeze(3) & 100k & \valerr{0.882}{0.001}\\% & \valerr{0.881}{0.001} \\
         & & Freeze(4) & 100k & \valerr{0.869}{0.002}\\% & \valerr{0.869}{0.002} \\
         & & Classifier Only & 100k & \valerr{0.790}{0.002}\\% & \valerr{0.790}{0.002} \\
         \noalign{\smallskip}
         \hline
         \noalign{\smallskip}
         & & & 1k & \valerr{0.794}{0.005}\\% & \valerr{0.798}{0.005} \\
         & & & 2k & \valerr{0.821}{0.002}\\% & \valerr{0.823}{0.002} \\
         & & & 3k & \valerr{0.824}{0.006}\\% & \valerr{0.826}{0.006} \\
         & & & 5k & \valerr{0.839}{0.003}\\% & \valerr{0.841}{0.002} \\ 
         & & & 7k & \valerr{0.847}{0.002}\\% & \valerr{0.848}{0.002} \\
         & & & 10k & \valerr{0.855}{0.002}\\% & \valerr{0.856}{0.001} \\
         ResNet18 & Pre-trained & All Weights & 15k & \valerr{0.862}{0.002}\\% & \valerr{0.862}{0.002} \\
         & & & 20k & \valerr{0.864}{0.002}\\% & \valerr{0.865}{0.001} \\
         & & & 30k & \valerr{0.872}{0.002}\\% & \valerr{0.872}{0.002} \\
         & & & 40k & \valerr{0.880}{0.002}\\% & \valerr{0.879}{0.002} \\
         & & & 50k & \valerr{0.880}{0.002}\\% & \valerr{0.880}{0.002} \\
         & & & 75k & \valerr{0.890}{0.002}\\% & \valerr{0.889}{0.002} \\
         & & & 100k & \valerr{0.896}{0.002}\\% & \valerr{0.895}{0.002} \\
         \noalign{\smallskip}
         \hline \hline
    \end{tabular}
    \label{tab:tfresults}
\end{table*}

The top section of Table~\ref{tab:tfresults} shows the comparison between the networks that had different numbers of weights free for fine-tuning when trained using the full sample of 100,000 interactions. As expected, the performance is best when allowing more weights to be fine-tuned since it gives the network more degrees of freedom to perform the classification. Within statistical uncertainties, the results from the All Weights and Freeze(1) categories are the same, which is to be expected since there are very few parameters in the first convolutional layer of the ResNet18. The best F1 score is hence reported as \valerr{0.896}{0.002} for fine tuning all of the network parameters. The fact that fine tuning the initial convolutional layer has very little effect on the CNN performance suggests that even though it was trained on photographic images, it is extracting generic features that are applicable to the LArTPC images. It is notable that training only the classifier weights still obtains a F1 score of \valerr{0.790}{0.002}, which, when compared to the results in Table~\ref{tab:riresults}, outperforms training the Kaiming-initialised network with fewer than 20k images. Furthermore, the addition of only the Layer 4 weights is sufficient to yield an F1 score of \valerr{0.869}{0.002}, out-performing the Kaiming-initialised network with the full 100k images.

\subsection{Comparison of Transfer Learning and Random Initialisation}
\label{sec:final_results}

\begin{figure*}%[tbh]
    \centering
    \includegraphics[width=0.75\textwidth]{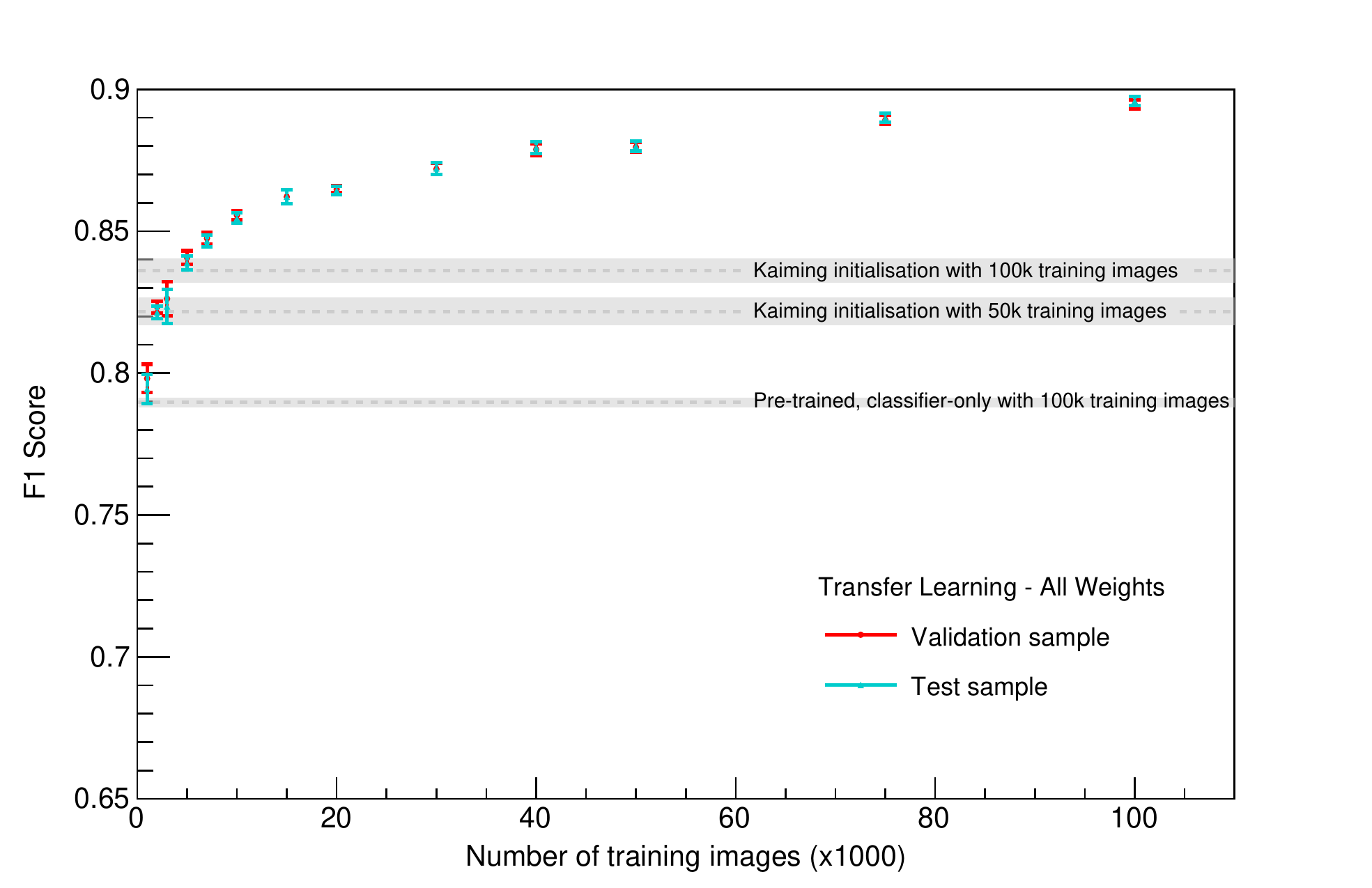}
    \caption{The test sample (cyan) and validation sample (red) F1 scores as a function of the number of images used to train the all-weights transfer learning networks. The three grey bands provide a visual guide to compare the performance to: Kaiming-initialised networks when trained with 100,000 (top) and 50,000 (middle) images, and transfer learning using 100,000 images in the classifier-only case (bottom).}
    \label{fig:val_f1score}
\end{figure*}

The F1 score measured for the transfer learning all weights case as a function of the number of training images is shown in the bottom section of Table~\ref{tab:tfresults}. Even when trained with 1k events it outperforms the classifier only case with 100k training examples. The results are shown graphically and compared to the randomly-initialised Kaiming networks for 50k and 100k training samples in Fig.~\ref{fig:val_f1score}. It shows that, in the case of fine tuning all of the network weights (cyan points for testing sample, red points for validation sample), the performance exceeds the Kaiming-initialised ResNet18 trained on 100k (50k) images using only 7k (5k) images. This is a powerful demonstration of the use of transfer learning even when a reasonably large training sample of 100k events is available to train a randomly-initialised CNN. Using all 100k events in the transfer learning case improves the F1 score from \valerr{0.836}{0.004} to \valerr{0.896}{0.002} compared to training the network from scratch with 100k events. The validation F1 score is shown to demonstrate that the network was able to generalise well to the test sample.

\begin{figure}%[tbh]
    \includegraphics[width=0.47\textwidth]{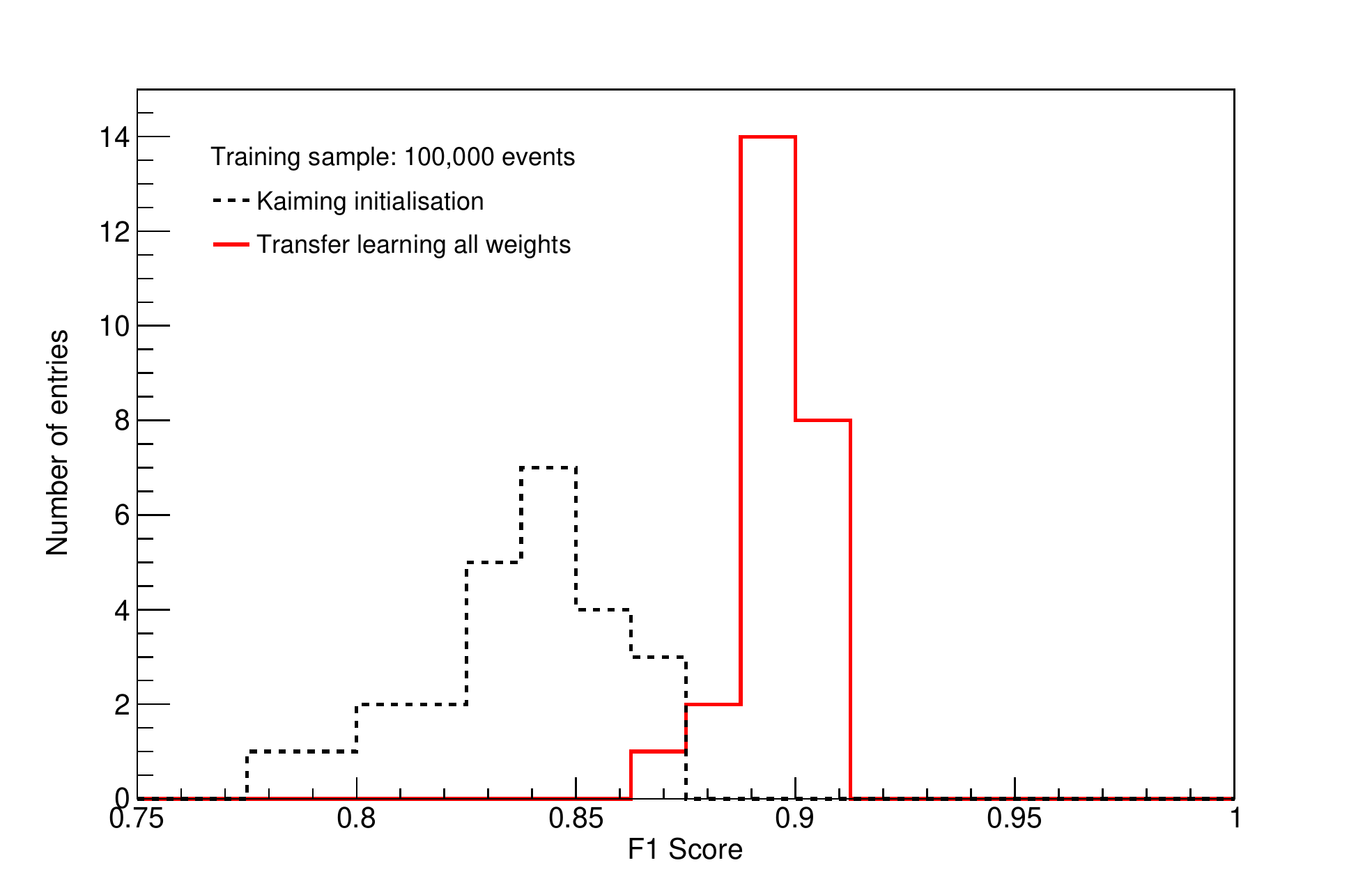}
    \caption{F1 score distributions from the ensembles of 25 trained networks using the transfer learning all weights method (red) and the Kaiming-initialised networks (black). In both cases the networks were trained using the same 100,000 events.}
    \label{fig:f1_score_dist}
\end{figure}

Figure~\ref{fig:f1_score_dist} shows the distribution of F1 scores from each of the 25 trained networks in the ensemble for the Kaiming-initialised (black) and the transfer learning all weights (red) cases, when trained using 100k interactions. The higher stability of the transfer learning case is shown clearly by the narrower distribution of F1 scores.

\begin{figure*}
    \centering
    \includegraphics[width=0.75\textwidth]{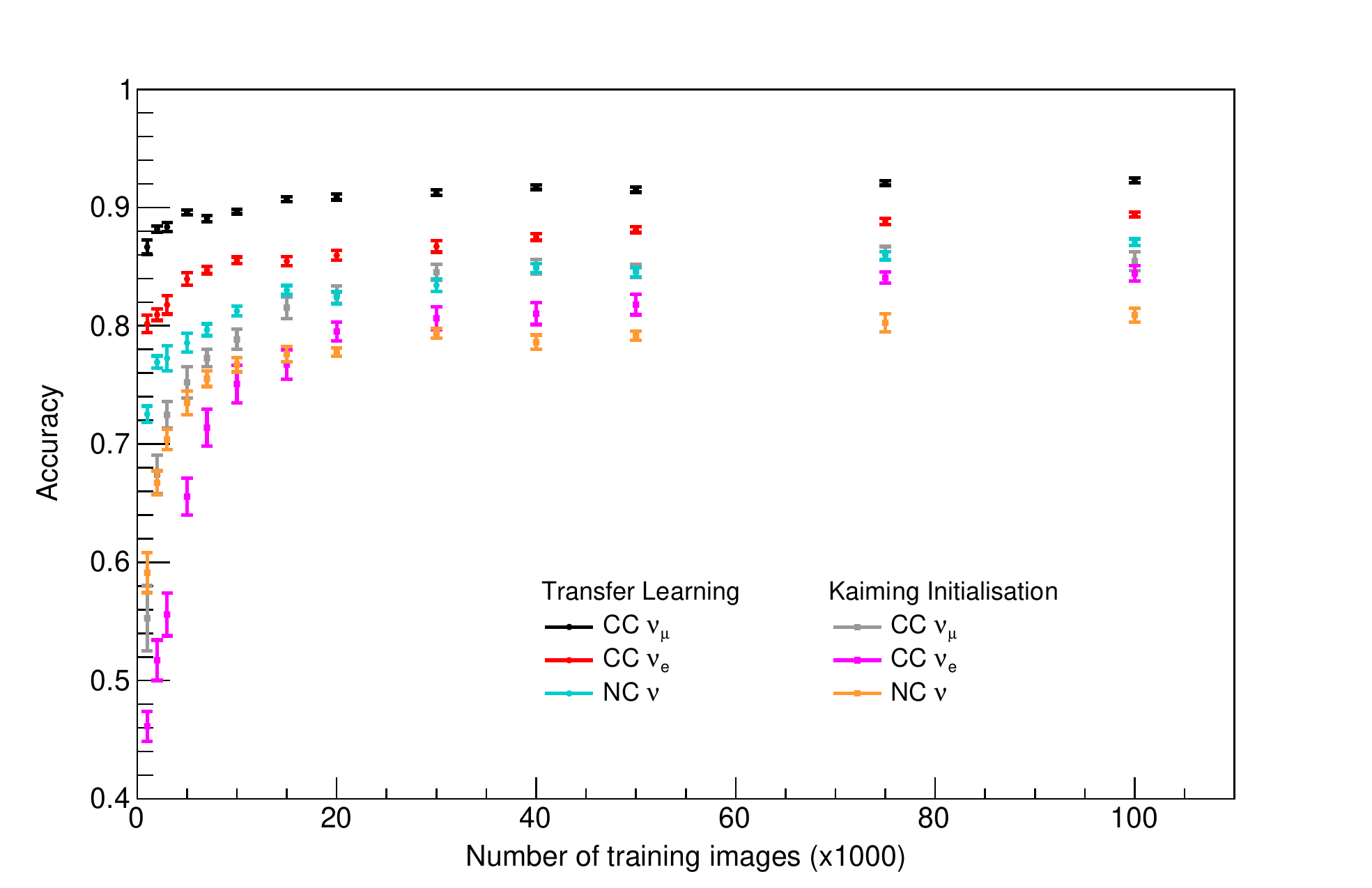}
    \caption{The accuracy for the three classes (\ccnumu, \ccnue{} and NC) for the transfer learning all weights and Kaiming-initialised networks. The error bars show the uncertainty on the mean accuracy from the ensembles of 25 trained networks for each point.}
    \label{fig:class_acc_tf_k}
\end{figure*}

Figure~\ref{fig:class_acc_tf_k} shows the class accuracy\footnote{Typically called efficiency in high energy physics contexts.} for the three classes: \ccnumu, \ccnue{} and NC, for the transfer learning all weights and Kaiming-initialised networks. The accuracy for each class in the transfer learning case exceeds the corresponding class performance using the Kaiming initialisation. It demonstrates that the improvements from transfer learning come from improvement in all three classes.

Table~\ref{tab:confusion_matrices} shows two example confusion matrices from the Kaiming-initialised and transfer learning all weights networks trained on 100k images. In both cases, the network with the best F1 score of the 25 networks in the ensemble was chosen for presentation. The diagonal terms in these matrices show the number of correctly classified events, and the off-diagonal terms show the number of events wrongly classified as either of the other true classes. It can be seen that the transfer learning all weights network shows more correctly classified events and fewer incorrectly classified events for each of the true classes.

\begin{table}[]
    \caption{Confusion matrices showing the number of correctly and incorrectly classified events for each true class. Results are shown for the network with the highest F1 score in the ensembles with 100,000 training images for Kaiming (top) and transfer learning all weights (bottom) cases.}
    \label{tab:confusion_matrices}
    
    \centering
    \begin{tabular}{cc|ccc}
      & & \multicolumn{3}{c}{Predicted} \\
      & & \ccnumu & \ccnue & NC \\ 
      \noalign{\smallskip}
      \hline
      \noalign{\smallskip}
      \multirow{3}{*}{True} & \ccnumu & 5839 & 386 & 487 \\
      & \ccnue & 276 & 5586 & 712 \\ 
      & NC & 337 & 624 & 5753 \\
      \noalign{\smallskip}
      \multicolumn{5}{c}{Kaiming initialisation}
    \end{tabular}
    
    \vspace{11pt}
    
    \begin{tabular}{cc|ccc}
      & & \multicolumn{3}{c}{Predicted} \\
      & & \ccnumu & \ccnue & NC \\ 
      \noalign{\smallskip}
      \hline
      \noalign{\smallskip}
      \multirow{3}{*}{True} & \ccnumu & 6262 & 163 & 287 \\
      & \ccnue & 111 & 6160 & 303 \\ 
      & NC & 228 & 367 & 6119 \\
      \noalign{\smallskip}
      \multicolumn{5}{c}{Transfer learning all weights}
    \end{tabular}
    
\end{table}

\subsection{Comparison of classification bias with neutrino energy}
\label{sec:bias_focused}

\begin{figure*}%[tbh]
    \centering
    \includegraphics[width=1.0\textwidth]{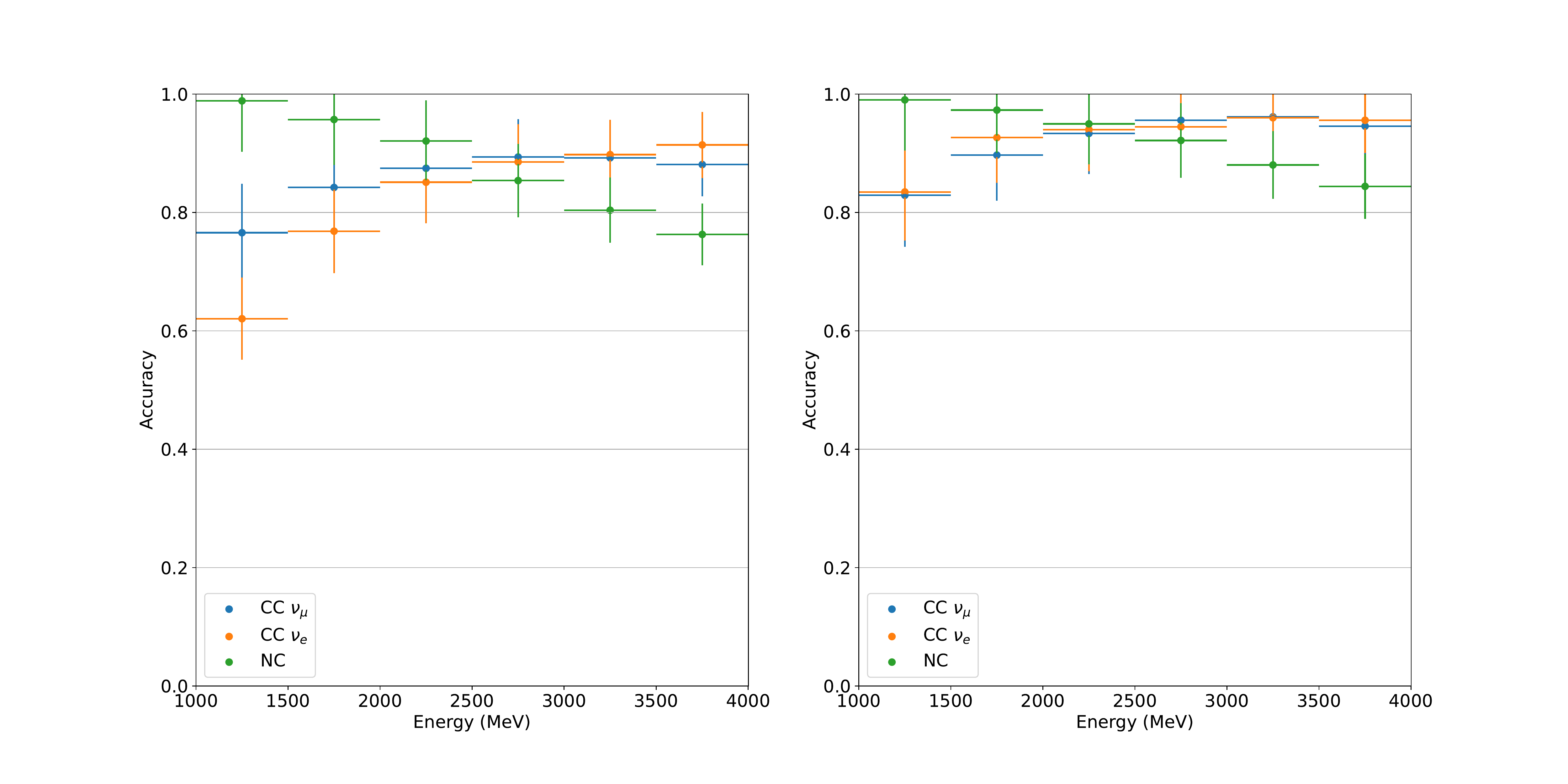}
    \caption{The classification accuracy for each interaction type as a function of the true neutrino energy for the Kaiming-initialised (left) and transfer learning all weights (right) cases for networks trained with 100,000 images.}
    \label{fig:bias_energy}
\end{figure*}
To compare potential biases in classification performance as a function of true neutrino energy we consider the most performant networks, according to F1 score, from the ensemble of 25 networks trained with 100k events for each of the Kaiming-initialised and transfer learning all weights cases.

Classification accuracy (the fraction of correctly classified events for a given true interaction type) on the 20k event test sample is shown in Fig.~\ref{fig:bias_energy}. It is evident that the classification performance does vary with energy, and the pattern of variation is similar for both the Kaiming-initialised network and the transfer learned network. Performance in the two charged-current classes is reduced at lower energies, but the magnitude of the bias is notably less in the transfer learned case, with charged-current performance more nearly equivalent in each energy bin. For neutral current interactions we see a reduction in performance as energy increases, but once again the network trained via transfer learning shows less bias.

\subsection{Comparison of classification bias with training sample size}
\label{sec:bias_focused}

\begin{figure*}%[tbh]
    \centering
    \includegraphics[width=1.0\textwidth]{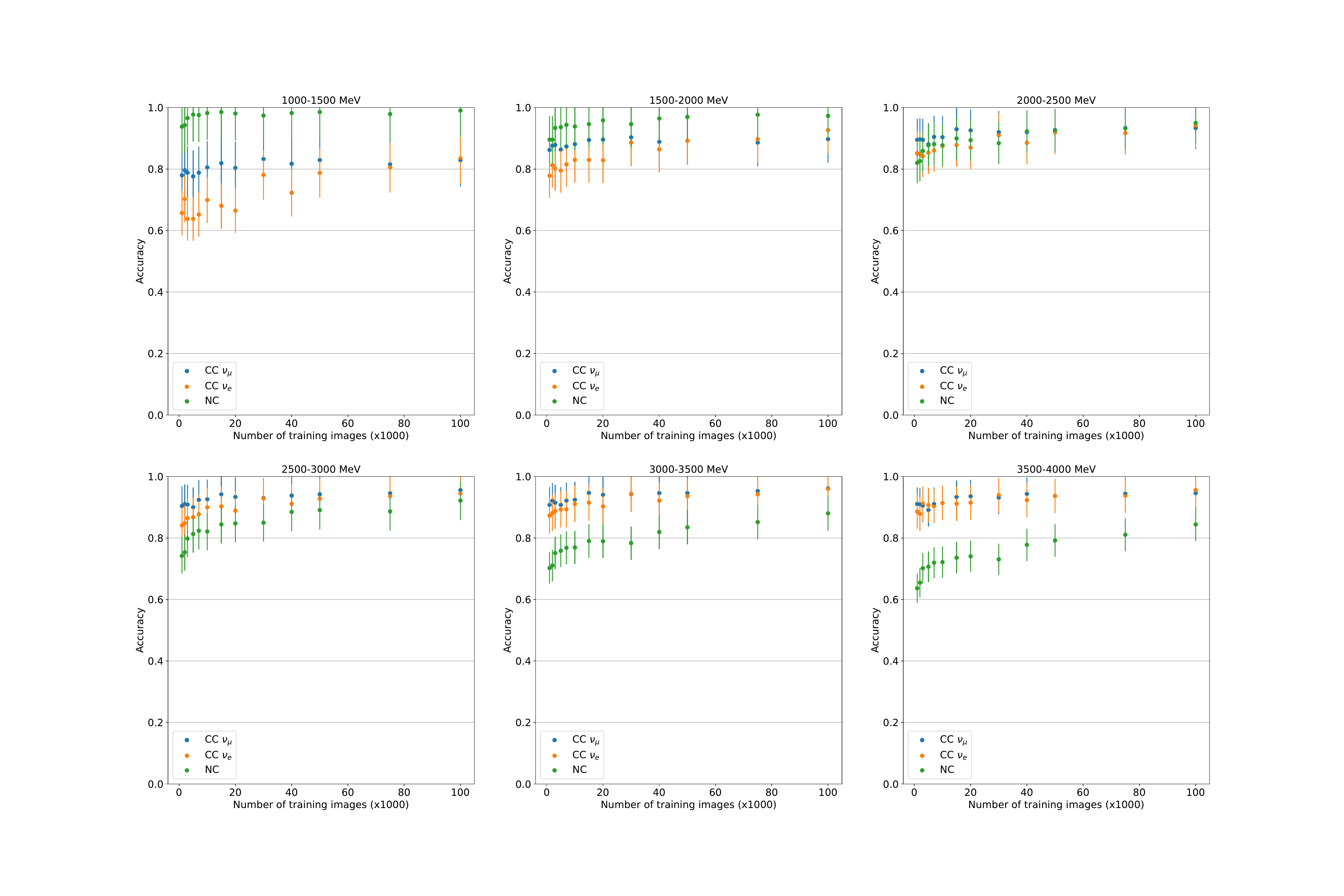}
    \caption{The classification accuracy for each interaction type as a function of the training sample size for transfer learning all weights cases.}
    \label{fig:bias_sample_size}
\end{figure*}

Given one of the potential benefits of transfer learning is the ability to train on fewer events, it is worthwhile to explore how the number of events affects classification performance at the level of interaction type. To compare potential biases in classification performance as a function of the training sample size we consider the most performant network within the ensemble of transfer learned networks for each size of training sample,

 First, it can be observed in Fig.~\ref{fig:bias_sample_size} that the same pattern of behaviour is evident across all sample sizes, that is, charged-current classification accuracy improves as energy increases, while neutral current classification accuracy is reduced. However, it is also evident that the number of events introduces its own biases for the lowest training sample sizes. In particular, it can be seen that the magnitude of the reduction in classification accuracy for sample sizes below approximately 30k events is sensitive to the particular true interaction type. Above this sample size, the improvement in classification accuracy is typically small, and similar for each interaction type.
 
Therefore, while the overall F1 score for transfer learning when training with only 7k events was broadly equivalent to that of the 100k event sample Kaiming-initialised network, it is clear that to achieve more balanced classification accuracy across interaction types and over a wide range of energies one should be cautious about pushing training sample size reduction too far without considering mitigating steps (for example, re-balancing the representation of each interaction type in the training sample). Though not presented here, such biases are, of course, also evident in the Kaiming-initialised networks, so this is not a unique feature of transfer learning.

\section{Discussion}\label{sec:discussion}
A systematic study comparing the performance of event classification in a DUNE-like LArTPC detector using standard CNN training methods and transfer learning was performed. It was found that the first convolutional layer of the ImageNet pre-trained ResNet18 extracted generic features that were applicable to LArTPC event images.

We have demonstrated that transfer learning can significantly outperform training randomly-initialised CNNs in the context of classifying neutrino interactions. We have also demonstrated that transfer learned networks exhibit reduced biases relative to networks trained from randomly initialised weights. Fine tuning a pre-trained ResNet18 with only 7k images gave a better F1 score than a Kaiming-initialised ResNet18 trained on 100k images, though evidence of classification biases at such low sample sizes, whatever the training method, indicates caution is required in the use of very small samples. The results presented here demonstrate the potential of the transfer learning method as a way to obtain very good CNN performance with a relatively small number of training images.

%% EPJC format
%\subsection*{Acknowledgments}
%The authors thank Prof. A.~Aurisano for the provision of computing resources at the University of Cincinnati.

% RevTex format
\begin{acknowledgments}
The authors thank Prof. A.~Aurisano for the provision of computing resources at the University of Cincinnati, Dr S.~Alonso~Monsalve for his helpful comments on the manuscript, and Dr A.~Basharina-Freshville for her work on a previous incarnation of the detector simulation.
\end{acknowledgments}
\vfill\null

\bibliography{mainbib}% Produces the bibliography via BibTeX.

\begin{thebibliography}{22}%
\makeatletter
\providecommand \@ifxundefined [1]{%
 \@ifx{#1\undefined}
}%
\providecommand \@ifnum [1]{%
 \ifnum #1\expandafter \@firstoftwo
 \else \expandafter \@secondoftwo
 \fi
}%
\providecommand \@ifx [1]{%
 \ifx #1\expandafter \@firstoftwo
 \else \expandafter \@secondoftwo
 \fi
}%
\providecommand \natexlab [1]{#1}%
\providecommand \enquote  [1]{``#1''}%
\providecommand \bibnamefont  [1]{#1}%
\providecommand \bibfnamefont [1]{#1}%
\providecommand \citenamefont [1]{#1}%
\providecommand \href@noop [0]{\@secondoftwo}%
\providecommand \href [0]{\begingroup \@sanitize@url \@href}%
\providecommand \@href[1]{\@@startlink{#1}\@@href}%
\providecommand \@@href[1]{\endgroup#1\@@endlink}%
\providecommand \@sanitize@url [0]{\catcode `\\12\catcode `\$12\catcode
  `\&12\catcode `\#12\catcode `\^12\catcode `\_12\catcode `\%12\relax}%
\providecommand \@@startlink[1]{}%
\providecommand \@@endlink[0]{}%
\providecommand \url  [0]{\begingroup\@sanitize@url \@url }%
\providecommand \@url [1]{\endgroup\@href {#1}{\urlprefix }}%
\providecommand \urlprefix  [0]{URL }%
\providecommand \Eprint [0]{\href }%
\providecommand \doibase [0]{https://doi.org/}%
\providecommand \selectlanguage [0]{\@gobble}%
\providecommand \bibinfo  [0]{\@secondoftwo}%
\providecommand \bibfield  [0]{\@secondoftwo}%
\providecommand \translation [1]{[#1]}%
\providecommand \BibitemOpen [0]{}%
\providecommand \bibitemStop [0]{}%
\providecommand \bibitemNoStop [0]{.\EOS\space}%
\providecommand \EOS [0]{\spacefactor3000\relax}%
\providecommand \BibitemShut  [1]{\csname bibitem#1\endcsname}%
\let\auto@bib@innerbib\@empty
%</preamble>
\bibitem [{\citenamefont {Psihas}\ \emph {et~al.}(2020)\citenamefont {Psihas},
  \citenamefont {Groh}, \citenamefont {Tunnell},\ and\ \citenamefont
  {Warburton}}]{Psihas2020}%
  \BibitemOpen
  \bibfield  {author} {\bibinfo {author} {\bibfnamefont {F.}~\bibnamefont
  {Psihas}}, \bibinfo {author} {\bibfnamefont {M.}~\bibnamefont {Groh}},
  \bibinfo {author} {\bibfnamefont {C.}~\bibnamefont {Tunnell}},\ and\ \bibinfo
  {author} {\bibfnamefont {K.}~\bibnamefont {Warburton}},\ }\bibfield  {title}
  {\bibinfo {title} {A review on machine learning for neutrino experiments},\
  }\href {https://doi.org/10.1142/s0217751x20430058} {\bibfield  {journal}
  {\bibinfo  {journal} {International Journal of Modern Physics A}\ }\textbf
  {\bibinfo {volume} {35}},\ \bibinfo {pages} {2043005} (\bibinfo {year}
  {2020})}\BibitemShut {NoStop}%
\bibitem [{\citenamefont {Radovic}\ \emph {et~al.}(2018)\citenamefont
  {Radovic}, \citenamefont {Williams}, \citenamefont {Rousseau}, \citenamefont
  {Kagan}, \citenamefont {Bonacorsi}, \citenamefont {Himmel}, \citenamefont
  {Aurisano}, \citenamefont {Terao},\ and\ \citenamefont
  {Wongjirad}}]{hepReview}%
  \BibitemOpen
  \bibfield  {author} {\bibinfo {author} {\bibfnamefont {A.}~\bibnamefont
  {Radovic}}, \bibinfo {author} {\bibfnamefont {M.}~\bibnamefont {Williams}},
  \bibinfo {author} {\bibfnamefont {D.}~\bibnamefont {Rousseau}}, \bibinfo
  {author} {\bibfnamefont {M.}~\bibnamefont {Kagan}}, \bibinfo {author}
  {\bibfnamefont {D.}~\bibnamefont {Bonacorsi}}, \bibinfo {author}
  {\bibfnamefont {A.}~\bibnamefont {Himmel}}, \bibinfo {author} {\bibfnamefont
  {A.}~\bibnamefont {Aurisano}}, \bibinfo {author} {\bibfnamefont
  {K.}~\bibnamefont {Terao}},\ and\ \bibinfo {author} {\bibfnamefont
  {T.}~\bibnamefont {Wongjirad}},\ }\bibfield  {title} {\bibinfo {title}
  {Machine learning at the energy and intensity frontiers of particle
  physics},\ }\href {https://doi.org/10.1038/s41586-018-0361-2} {\bibfield
  {journal} {\bibinfo  {journal} {Nature}\ }\textbf {\bibinfo {volume} {560}},\
  \bibinfo {pages} {41} (\bibinfo {year} {2018})}\BibitemShut {NoStop}%
\bibitem [{\citenamefont {Abi}\ \emph {et~al.}(2020{\natexlab{a}})\citenamefont
  {Abi} \emph {et~al.}}]{DUNE:2020gpm}%
  \BibitemOpen
  \bibfield  {author} {\bibinfo {author} {\bibfnamefont {B.}~\bibnamefont
  {Abi}} \emph {et~al.} (\bibinfo {collaboration} {DUNE}),\ }\bibfield  {title}
  {\bibinfo {title} {{Neutrino interaction classification with a convolutional
  neural network in the DUNE far detector}},\ }\href
  {https://doi.org/10.1103/PhysRevD.102.092003} {\bibfield  {journal} {\bibinfo
   {journal} {Phys. Rev. D}\ }\textbf {\bibinfo {volume} {102}},\ \bibinfo
  {pages} {092003} (\bibinfo {year} {2020}{\natexlab{a}})},\ \Eprint
  {https://arxiv.org/abs/2006.15052} {arXiv:2006.15052 [physics.ins-det]}
  \BibitemShut {NoStop}%
\bibitem [{\citenamefont {Alonso~Monsalve}(2020)}]{AlonsoMonsalve:2751646}%
  \BibitemOpen
  \bibfield  {author} {\bibinfo {author} {\bibfnamefont {S.}~\bibnamefont
  {Alonso~Monsalve}},\ }\emph {\bibinfo {title} {{Novel usage of deep learning
  and high-performance computing in long-baseline neutrino oscillation
  experiments}}},\ \href {https://cds.cern.ch/record/2751646} {Ph.D. thesis},\
  \bibinfo  {school} {Universidad Carlos III de Madrid} (\bibinfo {year}
  {2020})\BibitemShut {NoStop}%
\bibitem [{\citenamefont {Bloom}\ \emph {et~al.}(2022)\citenamefont {Bloom},
  \citenamefont {Boisvert}, \citenamefont {Britzger}, \citenamefont {Buuck},
  \citenamefont {Eichhorn}, \citenamefont {Headley}, \citenamefont
  {Lohwasser},\ and\ \citenamefont {Merkel}}]{climateAndHEP}%
  \BibitemOpen
  \bibfield  {author} {\bibinfo {author} {\bibfnamefont {K.}~\bibnamefont
  {Bloom}}, \bibinfo {author} {\bibfnamefont {V.}~\bibnamefont {Boisvert}},
  \bibinfo {author} {\bibfnamefont {D.}~\bibnamefont {Britzger}}, \bibinfo
  {author} {\bibfnamefont {M.}~\bibnamefont {Buuck}}, \bibinfo {author}
  {\bibfnamefont {A.}~\bibnamefont {Eichhorn}}, \bibinfo {author}
  {\bibfnamefont {M.}~\bibnamefont {Headley}}, \bibinfo {author} {\bibfnamefont
  {K.}~\bibnamefont {Lohwasser}},\ and\ \bibinfo {author} {\bibfnamefont
  {P.}~\bibnamefont {Merkel}},\ }\href
  {https://doi.org/10.48550/ARXIV.2203.12389} {\bibinfo {title} {Climate
  impacts of particle physics}} (\bibinfo {year} {2022}),\ \Eprint
  {https://arxiv.org/abs/arXiv:2203.12389} {arXiv:2203.12389} \BibitemShut
  {NoStop}%
\bibitem [{\citenamefont {Calafiura}\ \emph {et~al.}(2022)\citenamefont
  {Calafiura}, \citenamefont {Rousseau},\ and\ \citenamefont
  {Terao}}]{hepReview2}%
  \BibitemOpen
  \bibfield  {author} {\bibinfo {author} {\bibfnamefont {P.}~\bibnamefont
  {Calafiura}}, \bibinfo {author} {\bibfnamefont {D.}~\bibnamefont
  {Rousseau}},\ and\ \bibinfo {author} {\bibfnamefont {K.}~\bibnamefont
  {Terao}},\ }\href {https://doi.org/10.1142/12200} {\emph {\bibinfo {title}
  {Artificial Intelligence for High Energy Physics}}}\ (\bibinfo  {publisher}
  {World Scientific},\ \bibinfo {address} {Singapore},\ \bibinfo {year}
  {2022})\ \Eprint
  {https://arxiv.org/abs/https://worldscientific.com/doi/pdf/10.1142/12200}
  {https://worldscientific.com/doi/pdf/10.1142/12200} \BibitemShut {NoStop}%
\bibitem [{\citenamefont {Domin{\'{e}}}\ and\ \citenamefont
  {Terao}(2020)}]{Domin2020}%
  \BibitemOpen
  \bibfield  {author} {\bibinfo {author} {\bibfnamefont {L.}~\bibnamefont
  {Domin{\'{e}}}}\ and\ \bibinfo {author} {\bibfnamefont {K.}~\bibnamefont
  {Terao}},\ }\bibfield  {title} {\bibinfo {title} {Scalable deep convolutional
  neural networks for sparse, locally dense liquid argon time projection
  chamber data},\ }\bibfield  {journal} {\bibinfo  {journal} {Physical Review
  D}\ }\textbf {\bibinfo {volume} {102}},\ \href
  {https://doi.org/10.1103/physrevd.102.012005} {10.1103/physrevd.102.012005}
  (\bibinfo {year} {2020})\BibitemShut {NoStop}%
\bibitem [{\citenamefont {Bozinovski}\ and\ \citenamefont
  {Fulgosi}(1976)}]{transfer1}%
  \BibitemOpen
  \bibfield  {author} {\bibinfo {author} {\bibfnamefont {S.}~\bibnamefont
  {Bozinovski}}\ and\ \bibinfo {author} {\bibfnamefont {A.}~\bibnamefont
  {Fulgosi}},\ }\bibfield  {title} {\bibinfo {title} {The influence of pattern
  similarity and transfer learning upon the training of a base perceptron b2},\
  }in\ \href@noop {} {\emph {\bibinfo {booktitle} {Proceedings of Symposium
  Informatica}}}\ (\bibinfo {address} {Bled, Slovenia},\ \bibinfo {year}
  {1976})\ pp.\ \bibinfo {pages} {3--121--5},\ \bibinfo {note} {original in
  Croatian}\BibitemShut {NoStop}%
\bibitem [{\citenamefont {Bozinovski}(2020)}]{transfer2}%
  \BibitemOpen
  \bibfield  {author} {\bibinfo {author} {\bibfnamefont {S.}~\bibnamefont
  {Bozinovski}},\ }\bibfield  {title} {\bibinfo {title} {{Reminder of the first
  paper on transfer learning in neural networks, 1976}},\ }\href@noop {}
  {\bibfield  {journal} {\bibinfo  {journal} {Informatica}\ }\textbf {\bibinfo
  {volume} {44}},\ \bibinfo {pages} {291–302} (\bibinfo {year}
  {2020})}\BibitemShut {NoStop}%
\bibitem [{\citenamefont {Zhuang}\ \emph {et~al.}(2020)\citenamefont {Zhuang},
  \citenamefont {Qi}, \citenamefont {Duan}, \citenamefont {Xi}, \citenamefont
  {Zhu}, \citenamefont {Zhu}, \citenamefont {Xiong},\ and\ \citenamefont
  {He}}]{transferReview}%
  \BibitemOpen
  \bibfield  {author} {\bibinfo {author} {\bibfnamefont {F.}~\bibnamefont
  {Zhuang}}, \bibinfo {author} {\bibfnamefont {Z.}~\bibnamefont {Qi}}, \bibinfo
  {author} {\bibfnamefont {K.}~\bibnamefont {Duan}}, \bibinfo {author}
  {\bibfnamefont {D.}~\bibnamefont {Xi}}, \bibinfo {author} {\bibfnamefont
  {Y.}~\bibnamefont {Zhu}}, \bibinfo {author} {\bibfnamefont {H.}~\bibnamefont
  {Zhu}}, \bibinfo {author} {\bibfnamefont {H.}~\bibnamefont {Xiong}},\ and\
  \bibinfo {author} {\bibfnamefont {Q.}~\bibnamefont {He}},\ }\bibfield
  {title} {\bibinfo {title} {A comprehensive survey on transfer learning},\
  }\href {https://doi.org/10.1109/JPROC.2020.3004555} {\bibfield  {journal}
  {\bibinfo  {journal} {Proceedings of the IEEE}\ }\textbf {\bibinfo {volume}
  {PP}},\ \bibinfo {pages} {1} (\bibinfo {year} {2020})}\BibitemShut {NoStop}%
\bibitem [{\citenamefont {Kuchera}\ \emph {et~al.}(2019)\citenamefont
  {Kuchera}, \citenamefont {Ramanujan}, \citenamefont {Taylor}, \citenamefont
  {Strauss}, \citenamefont {Bazin}, \citenamefont {Bradt},\ and\ \citenamefont
  {Chen}}]{Kuchera:2018djs}%
  \BibitemOpen
  \bibfield  {author} {\bibinfo {author} {\bibfnamefont {M.~P.}\ \bibnamefont
  {Kuchera}}, \bibinfo {author} {\bibfnamefont {R.}~\bibnamefont {Ramanujan}},
  \bibinfo {author} {\bibfnamefont {J.~Z.}\ \bibnamefont {Taylor}}, \bibinfo
  {author} {\bibfnamefont {R.~R.}\ \bibnamefont {Strauss}}, \bibinfo {author}
  {\bibfnamefont {D.}~\bibnamefont {Bazin}}, \bibinfo {author} {\bibfnamefont
  {J.}~\bibnamefont {Bradt}},\ and\ \bibinfo {author} {\bibfnamefont
  {R.}~\bibnamefont {Chen}},\ }\bibfield  {title} {\bibinfo {title} {{Machine
  Learning Methods for Track Classification in the AT-TPC}},\ }\href
  {https://doi.org/10.1016/j.nima.2019.05.097} {\bibfield  {journal} {\bibinfo
  {journal} {Nucl. Instrum. Meth. A}\ }\textbf {\bibinfo {volume} {940}},\
  \bibinfo {pages} {156} (\bibinfo {year} {2019})},\ \Eprint
  {https://arxiv.org/abs/1810.10350} {arXiv:1810.10350 [cs.CV]} \BibitemShut
  {NoStop}%
\bibitem [{\citenamefont {Alam}\ \emph {et~al.}(2015)\citenamefont {Alam},
  \citenamefont {Andreopoulos}, \citenamefont {Athar}, \citenamefont {Bodek},
  \citenamefont {Christy}, \citenamefont {Coopersmith}, \citenamefont {Dennis},
  \citenamefont {Dytman}, \citenamefont {Gallagher}, \citenamefont {Geary},
  \citenamefont {Golan}, \citenamefont {Hatcher}, \citenamefont {Hoshina},
  \citenamefont {Liu}, \citenamefont {Mahn}, \citenamefont {Marshall},
  \citenamefont {Morrison}, \citenamefont {Nirkko}, \citenamefont {Nowak},
  \citenamefont {Perdue},\ and\ \citenamefont {Yarba}}]{Alam:2015nkk}%
  \BibitemOpen
  \bibfield  {author} {\bibinfo {author} {\bibfnamefont {M.}~\bibnamefont
  {Alam}}, \bibinfo {author} {\bibfnamefont {C.}~\bibnamefont {Andreopoulos}},
  \bibinfo {author} {\bibfnamefont {M.}~\bibnamefont {Athar}}, \bibinfo
  {author} {\bibfnamefont {A.}~\bibnamefont {Bodek}}, \bibinfo {author}
  {\bibfnamefont {E.}~\bibnamefont {Christy}}, \bibinfo {author} {\bibfnamefont
  {B.}~\bibnamefont {Coopersmith}}, \bibinfo {author} {\bibfnamefont
  {S.}~\bibnamefont {Dennis}}, \bibinfo {author} {\bibfnamefont
  {S.}~\bibnamefont {Dytman}}, \bibinfo {author} {\bibfnamefont
  {H.}~\bibnamefont {Gallagher}}, \bibinfo {author} {\bibfnamefont
  {N.}~\bibnamefont {Geary}}, \bibinfo {author} {\bibfnamefont
  {T.}~\bibnamefont {Golan}}, \bibinfo {author} {\bibfnamefont
  {R.}~\bibnamefont {Hatcher}}, \bibinfo {author} {\bibfnamefont
  {K.}~\bibnamefont {Hoshina}}, \bibinfo {author} {\bibfnamefont
  {J.}~\bibnamefont {Liu}}, \bibinfo {author} {\bibfnamefont {K.}~\bibnamefont
  {Mahn}}, \bibinfo {author} {\bibfnamefont {C.}~\bibnamefont {Marshall}},
  \bibinfo {author} {\bibfnamefont {J.}~\bibnamefont {Morrison}}, \bibinfo
  {author} {\bibfnamefont {M.}~\bibnamefont {Nirkko}}, \bibinfo {author}
  {\bibfnamefont {J.}~\bibnamefont {Nowak}}, \bibinfo {author} {\bibfnamefont
  {G.~N.}\ \bibnamefont {Perdue}},\ and\ \bibinfo {author} {\bibfnamefont
  {J.}~\bibnamefont {Yarba}},\ }\href
  {https://doi.org/10.48550/ARXIV.1512.06882} {\bibinfo {title} {Genie
  production release 2.10.0}} (\bibinfo {year} {2015})\BibitemShut {NoStop}%
\bibitem [{\citenamefont {Abi}\ \emph {et~al.}(2020{\natexlab{b}})\citenamefont
  {Abi} \emph {et~al.}}]{DUNE:2020jqi}%
  \BibitemOpen
  \bibfield  {author} {\bibinfo {author} {\bibfnamefont {B.}~\bibnamefont
  {Abi}} \emph {et~al.} (\bibinfo {collaboration} {DUNE}),\ }\bibfield  {title}
  {\bibinfo {title} {{Long-baseline neutrino oscillation physics potential of
  the DUNE experiment}},\ }\href
  {https://doi.org/10.1140/epjc/s10052-020-08456-z} {\bibfield  {journal}
  {\bibinfo  {journal} {Eur. Phys. J. C}\ }\textbf {\bibinfo {volume} {80}},\
  \bibinfo {pages} {978} (\bibinfo {year} {2020}{\natexlab{b}})},\ \Eprint
  {https://arxiv.org/abs/2006.16043} {arXiv:2006.16043 [hep-ex]} \BibitemShut
  {NoStop}%
\bibitem [{\citenamefont {Agostinelli}\ \emph {et~al.}(2003)\citenamefont
  {Agostinelli} \emph {et~al.}}]{Agostinelli:2002hh}%
  \BibitemOpen
  \bibfield  {author} {\bibinfo {author} {\bibfnamefont {S.}~\bibnamefont
  {Agostinelli}} \emph {et~al.} (\bibinfo {collaboration} {GEANT4}),\
  }\bibfield  {title} {\bibinfo {title} {{GEANT4: A Simulation toolkit}},\
  }\href {https://doi.org/10.1016/S0168-9002(03)01368-8} {\bibfield  {journal}
  {\bibinfo  {journal} {Nucl.\ Instrum.\ Meth.\ A}\ }\textbf {\bibinfo {volume}
  {506}},\ \bibinfo {pages} {250} (\bibinfo {year} {2003})}\BibitemShut
  {NoStop}%
%%CITATION = NUIMA,A506,250;%%
\bibitem [{\citenamefont {Abi}\ \emph {et~al.}(2020{\natexlab{c}})\citenamefont
  {Abi} \emph {et~al.}}]{DUNE:2020txw}%
  \BibitemOpen
  \bibfield  {author} {\bibinfo {author} {\bibfnamefont {B.}~\bibnamefont
  {Abi}} \emph {et~al.} (\bibinfo {collaboration} {DUNE}),\ }\bibfield  {title}
  {\bibinfo {title} {{Deep Underground Neutrino Experiment (DUNE), Far Detector
  Technical Design Report, Volume IV: Far Detector Single-phase Technology}},\
  }\href {https://doi.org/10.1088/1748-0221/15/08/T08010} {\bibfield  {journal}
  {\bibinfo  {journal} {JINST}\ }\textbf {\bibinfo {volume} {15}}\bibfield
  {number} {\bibinfo  {number} { (08)},\ \bibinfo {pages} {T08010}},\ }\Eprint
  {https://arxiv.org/abs/2002.03010} {arXiv:2002.03010 [physics.ins-det]}
  \BibitemShut {NoStop}%
\bibitem [{\citenamefont {Paszke}\ \emph {et~al.}(2019)\citenamefont {Paszke}
  \emph {et~al.}}]{NEURIPS2019_9015}%
  \BibitemOpen
  \bibfield  {author} {\bibinfo {author} {\bibfnamefont {A.}~\bibnamefont
  {Paszke}} \emph {et~al.},\ }\bibfield  {title} {\bibinfo {title} {Pytorch: An
  imperative style, high-performance deep learning library},\ }in\ \href
  {http://papers.neurips.cc/paper/9015-pytorch-an-imperative-style-high-performance-deep-learning-library.pdf}
  {\emph {\bibinfo {booktitle} {Advances in Neural Information Processing
  Systems 32}}},\ \bibinfo {editor} {edited by\ \bibinfo {editor}
  {\bibfnamefont {H.}~\bibnamefont {Wallach}}, \bibinfo {editor} {\bibfnamefont
  {H.}~\bibnamefont {Larochelle}}, \bibinfo {editor} {\bibfnamefont
  {A.}~\bibnamefont {Beygelzimer}}, \bibinfo {editor} {\bibfnamefont
  {F.}~\bibnamefont {d\textquotesingle Alch\'{e}-Buc}}, \bibinfo {editor}
  {\bibfnamefont {E.}~\bibnamefont {Fox}},\ and\ \bibinfo {editor}
  {\bibfnamefont {R.}~\bibnamefont {Garnett}}}\ (\bibinfo  {publisher} {Curran
  Associates, Inc.},\ \bibinfo {address} {Vancouver, Canada},\ \bibinfo {year}
  {2019})\ pp.\ \bibinfo {pages} {8024--8035}\BibitemShut {NoStop}%
\bibitem [{\citenamefont {He}\ \emph {et~al.}(2015{\natexlab{a}})\citenamefont
  {He}, \citenamefont {Zhang}, \citenamefont {Ren},\ and\ \citenamefont
  {Sun}}]{He-et-al-2015-deep}%
  \BibitemOpen
  \bibfield  {author} {\bibinfo {author} {\bibfnamefont {K.}~\bibnamefont
  {He}}, \bibinfo {author} {\bibfnamefont {X.}~\bibnamefont {Zhang}}, \bibinfo
  {author} {\bibfnamefont {S.}~\bibnamefont {Ren}},\ and\ \bibinfo {author}
  {\bibfnamefont {J.}~\bibnamefont {Sun}},\ }\bibfield  {title} {\bibinfo
  {title} {Deep residual learning for image recognition},\ }\href
  {http://arxiv.org/abs/1512.03385} {\bibfield  {journal} {\bibinfo  {journal}
  {CoRR}\ }\textbf {\bibinfo {volume} {abs/1512.03385}} (\bibinfo {year}
  {2015}{\natexlab{a}})},\ \Eprint {https://arxiv.org/abs/1512.03385}
  {arXiv:1512.03385} \BibitemShut {NoStop}%
\bibitem [{\citenamefont {Van~Rijsbergen}(1979)}]{f1score}%
  \BibitemOpen
  \bibfield  {author} {\bibinfo {author} {\bibfnamefont {C.~J.}\ \bibnamefont
  {Van~Rijsbergen}},\ }\href {http://www.dcs.gla.ac.uk/Keith/Preface.html}
  {\emph {\bibinfo {title} {Information Retrieval}}},\ \bibinfo {edition}
  {2nd}\ ed.\ (\bibinfo  {publisher} {Butterworths},\ \bibinfo {address}
  {London, United Kingdom},\ \bibinfo {year} {1979})\ \Eprint
  {https://arxiv.org/abs/http://www.dcs.gla.ac.uk/Keith/Preface.html}
  {http://www.dcs.gla.ac.uk/Keith/Preface.html} \BibitemShut {NoStop}%
\bibitem [{\citenamefont {He}\ \emph {et~al.}(2015{\natexlab{b}})\citenamefont
  {He}, \citenamefont {Zhang}, \citenamefont {Ren},\ and\ \citenamefont
  {Sun}}]{kaiming}%
  \BibitemOpen
  \bibfield  {author} {\bibinfo {author} {\bibfnamefont {K.}~\bibnamefont
  {He}}, \bibinfo {author} {\bibfnamefont {X.}~\bibnamefont {Zhang}}, \bibinfo
  {author} {\bibfnamefont {S.}~\bibnamefont {Ren}},\ and\ \bibinfo {author}
  {\bibfnamefont {J.}~\bibnamefont {Sun}},\ }\href
  {https://doi.org/10.48550/ARXIV.1502.01852} {\bibinfo {title} {Delving deep
  into rectifiers: Surpassing human-level performance on imagenet
  classification}} (\bibinfo {year} {2015}{\natexlab{b}})\BibitemShut {NoStop}%
\bibitem [{\citenamefont {Deng}\ \emph {et~al.}(2009)\citenamefont {Deng},
  \citenamefont {Dong}, \citenamefont {Socher}, \citenamefont {Li},
  \citenamefont {Li},\ and\ \citenamefont {Fei-Fei}}]{imagenet_cvpr09}%
  \BibitemOpen
  \bibfield  {author} {\bibinfo {author} {\bibfnamefont {J.}~\bibnamefont
  {Deng}}, \bibinfo {author} {\bibfnamefont {W.}~\bibnamefont {Dong}}, \bibinfo
  {author} {\bibfnamefont {R.}~\bibnamefont {Socher}}, \bibinfo {author}
  {\bibfnamefont {L.-J.}\ \bibnamefont {Li}}, \bibinfo {author} {\bibfnamefont
  {K.}~\bibnamefont {Li}},\ and\ \bibinfo {author} {\bibfnamefont
  {L.}~\bibnamefont {Fei-Fei}},\ }\bibfield  {title} {\bibinfo {title}
  {{ImageNet: A Large-Scale Hierarchical Image Database}},\ }in\ \href@noop {}
  {\emph {\bibinfo {booktitle} {CVPR09}}}\ (\bibinfo {year} {2009})\BibitemShut
  {NoStop}%
\bibitem [{\citenamefont {Zeiler}\ and\ \citenamefont {Fergus}(2013)}]{zeiler}%
  \BibitemOpen
  \bibfield  {author} {\bibinfo {author} {\bibfnamefont {M.~D.}\ \bibnamefont
  {Zeiler}}\ and\ \bibinfo {author} {\bibfnamefont {R.}~\bibnamefont
  {Fergus}},\ }\href {https://doi.org/10.48550/arXiv.1311.2901} {\bibinfo
  {title} {Visualizing and understanding convolutional networks}} (\bibinfo
  {year} {2013}),\ \Eprint {https://arxiv.org/abs/arXiv:1311.2901}
  {arXiv:1311.2901} \BibitemShut {NoStop}%
\bibitem [{Note1()}]{Note1}%
  \BibitemOpen
  \bibinfo {note} {Typically called efficiency in high energy physics
  contexts.}\BibitemShut {Stop}%
\end{thebibliography}%

\end{document}